\documentclass{moriond}

\def\Journal#1#2#3#4{{#1} {\bf #2}, #3 (#4)}

\def\PRD{{\em Phys. Rev.} D}

\def\be{\begin{equation}}
\def\ee{\end{equation}}
\def\bea{\begin{eqnarray}}
\def\eea{\end{eqnarray}}

\newcommand{\Photo}{\includegraphics[height=20mm]{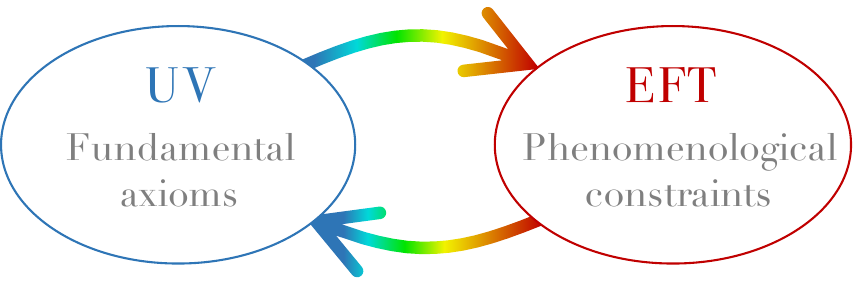}\vspace{-0.2cm}}

\begin{document}
\vspace*{4cm}
\title{Amplitude constraints on dark energy}

\author{ Scott Melville }

\address{Department of Physics and Astronomy, Queen Mary University of London, \\
Mile End Road, London E1 4NS, United Kingdom}

\maketitle\abstracts{
This talk gives a short introduction to the ``UV/EFT correspondence'', which uses \mbox{scattering} amplitudes to relate the Effective Field Theory (EFT) coefficients probed by low-energy \mbox{measurements} to properties of the underlying high-energy (UV) completion.
This includes \mbox{recent} ``positivity bounds'' on EFT coefficients, which are the low-energy signatures of \mbox{causality} and unitarity in the UV.
To illustrate their phenomenological impact, I apply these bounds to a simple EFT for dark energy and compare with recent cosmological observations.}

\paragraph{Introduction.}
%
Despite their unrivalled empirical success---describing a range of observations from the solar system to TeV colliders---General Relativity (GR) and the Standard Model (SM) cannot be a complete description of our Universe.
Observationally, there are growing \mbox{discrepancies} at both ends of this range, with colliders beginning to probe deviations from SM predictions on small length scales and cosmological surveys signalling new physics (e.g. dark matter/dark energy) on large length scales.
Theoretically, the breakdown of GR's predictivity in the UV and fine-tuning problems with the cosmological constant in the IR also point towards new physics on both small and large scales.

For experiments searching for small deviations from GR or the SM at these frontiers, \mbox{Effective} Field Theory (EFT) is increasingly the framework of choice, in which the \mbox{phenomenological} \mbox{effects} of new small-scale physics is captured by systematically adding corrections to the \mbox{theory} with undetermined coefficients that can be readily compared with large-scale data. 
While \mbox{pragmatically} useful, this approach presents two interesting issues. First, there is no guarantee that all values of these EFT coefficients are physical---we may be fitting our data with values that are secretly inconsistent (i.e. values that could never actually arise from any healthy small-scale physics). Second, what high-energy (UV) information about the underlying fundamental physics can we actually extract from a measurement of these EFT coefficients?

In this talk, I will give an overview of the emerging ``UV/EFT Correspondence''---explicit connections between properties of the UV physics and regions of the EFT parameter space which seek to address these issues---and discuss how this impacts a particular EFT of dark energy. \vspace{-0.17cm}

\paragraph{UV/EFT Correspondence.}
%
The main idea is to use the $2 \to 2$ scattering amplitude as a bridge between EFT data and UV matrix elements.
Suppose an initial 2-particle state $| p_1 p_2 \rangle$ evolves into $(1 + i \hat{T} ) | p_1  p_2 \rangle$, where $\hat{T}$ accounts for the interactions between the particles (which can change their momenta/state).
Thanks to Poincar\'{e} invariance, the $2 \to 2$ transition matrix,
\bea
\langle p_3 p_4 | \hat{T} | p_1 p_2 \rangle = A (s,t) \, (2 \pi )^4 \delta^4 \left( p_1 + p_2 - p_3 - p_4 \right)
\eea 
contains an amplitude $A$ which depends only on $s = -(p_1 + p_2)^2$ and $t = -(p_1 - p_3)^2$.\footnote{
	A third variable, $u=- (p_1 - p_4)^2$, is related to the others by $s+t+u =\sum_{a=1}^4 m_a^2$ since $p_a^2 = - m_a^2$. 
}
The EFT expansion corresponds to a Taylor series expansion of the part of $A$ mediated by heavy fields, 
\bea
 A (s,t) &=& \mbox{light physics} + \mbox{heavy physics} \qquad \left( \mathrm{e.g.} \;\; \frac{m^2}{m^2-s} + \frac{M^2}{M^2- s }  \right)  \; , \nonumber \\ 
\Rightarrow \;\;  A_{\rm EFT} (s,t) &=& \mbox{light physics} + \sum_{a,b} c_{ab} \, s^a t^b  \qquad  \quad \;\; \left( \mathrm{e.g.} \;\; \frac{m^2}{m^2-s} + \, \sum_{a} \frac{s^a}{M^a} \, \right)  \, .  \label{eqn:eft_exp}  
\eea
The goal is to extract from the $c_{ab}$ coefficients (which can be measured in low-energy experiments) concrete properties of the underlying heavy fields. Mathematically, this amounts to asking: how much can you say about a function if you only know its first few Taylor series coefficients? 

Amplitudes are well-suited for this problem because \emph{causal} interactions lead to \emph{analytic} $A(s,t)$ in the complex $s$-plane, modulo branch cuts from the exchange of on-shell particles.
Cauchy's theorem can then express $A(s,t)$ as a contour integral around these branch cuts, and the EFT expansion in (\ref{eqn:eft_exp}) corresponds to splitting this into two regions: $\int_0^{M^2} ds$ and $\int_{M^2}^{\infty}ds$.
For the latter, while we may not know the precise high-energy fields at $s > M^2$, since the underlying theory is Poincar\'{e} invariant it must have operators $\hat{P}_\mu$ and $\hat{J}$ that generate translations and rotations in the scattering plane. The second key property of amplitudes is that \emph{unitary} time evolution fixes the branch cut discontinuity in terms of a matrix element of these generators,\footnote{
The prime denotes that an overall momentum-conserving $\delta$-function has been removed.
}
\bea
\int_{M^2}^{\infty} \frac{ds}{\pi s} \frac{ \mathrm{Im} \, A (s, t) }{s^{a}} = \int_{M^2}^{\infty} \frac{ds}{2 \pi s} \, \langle p_1 p_2 | \hat{T}^{\dagger}  \frac{ \cos \left( \hat{J} \theta_{\hat{P}} (t) \right) }{ \hat{P}^{2a} }  \hat{T} | p_1 p_2 \rangle'    \equiv \left\langle  \frac{ \cos \left( \hat{J} \theta_{\hat{P}} (t) \right) }{ \hat{P}^{2a} }  \right\rangle_{P^2 > M^2} \, ,
\label{eqn:unitary}
\eea
where $\cos \theta_{\hat{P}} (t) = 1 + 2 t/( \hat{P}^2 - \sum_{a=1}^4 m_a^2 )$. 
Altogether, causality and unitarity then imply that,\footnote{
In order to ensure convergence of these integrals, locality (the Froissart bound) is also implicitly assumed.
}$\,$\footnote{
	To simplify the presentation of~(\ref{eqn:cab_int_rep}), $\mathcal{O}(m/M)$ corrections have been neglected and I assume $a \geq 2$.
} 
\bea
 c_{ab} = \frac{\partial_t^b}{b!} \frac{\partial_s^a}{a!}\left[  \left\langle \cos \left( \hat{J} \theta_{\hat{P}} (t) \right) \left( \frac{ 1 }{ \hat{P}^2 - s  }  +  \frac{ 1 }{ \hat{P}^2 - u } \right) \right\rangle_{P^2 > M^2}  \right]_{s=t=0}\; . 
\label{eqn:cab_int_rep}
\eea 
The left-hand-side is a low-energy EFT coefficient, which can be probed directly by experiment. The right-hand-side is a UV matrix element (an average over states with invariant mass $>M^2$), which is determined by properties of the underlying heavy fields.
For instance$\,$\cite{Adams:2006sv,deRham:2017avq},
\bea
  c_{20} = \left\langle \frac{2}{\hat{P}^4} \right\rangle_{P^2 > M^2}   \; ,
\qquad\qquad 2 M^2 c_{21} + 3 c_{20}   = \left\langle \frac{ 8 M^2 \hat{J}^2 + 6 ( \hat{P}^2 - M^2 ) }{\hat{P}^6} \right\rangle_{P^2 > M^2}      \; . \label{eqn:pos} 
\eea
Since these UV averages are positive, the corresponding $c_{ab}$ must satisfy certain ``positivity bounds''. For instance, regions of parameter space in which $c_{20} < 0$ are secretly unphysical because they would require high-energy states for which $\langle \hat{P}^{-4} \rangle < 0$ and hence violate unitarity.  \vspace{-0.8cm}

\paragraph{Dark Energy.}
%
One of the clearest indications of new physics in our Universe is its accelerated expansion. Explaining the observed rate within the SM is difficult (perhaps impossible).
So for an illustration of the above UV/EFT correspondence, I will consider a simple EFT which contains, in addition to the metric $g_{\mu\nu}$ of GR and the SM fields, a new scalar field $\phi$ which plays the role of dark energy on cosmological scales. 
In particular, I will focus on the EFT action,\footnote{
	(\ref{eqn:Horndeski}) is the quartic, shift-symmetric Horndeski theory: the most general EFT of these fields at this derivative order which is invariant under diffeomorphisms, $\phi \to - \phi$, and a (weakly broken) Galileon symmetry, $\phi \to \phi + c_\mu x^\mu$.}
\begin{equation}
	S = \int d^4 x \, \sqrt{-g} \left(  M_P^2 G_4 (X) R +  M_P G_4' (X) \frac{  \Phi^{\mu}_{\mu} \Phi^{\nu}_{\nu} - \Phi_{\mu}^{\nu} \Phi^{\mu}_{\nu} }{M^3}  + P (X) + \mathcal{L}_{\rm mat} \left( \psi, g_{\mu\nu} \right) \right) .
	\label{eqn:Horndeski}
\end{equation}
The first term is the usual Einstein-Hilbert action of GR but with the Planck mass rescaled by a free function $G_4 (X)$ of the dimensionless $X = \frac{-g^{\mu\nu} \nabla_\mu \phi \nabla_\nu \phi}{2 M_P M^3}$.
The second term is a particular scalar self-interaction, built from the matrix $\Phi_{\mu}^{\nu} = \nabla_\mu \nabla^\nu \phi$, which ensures that the theory is free from Ostrogradski-type instabilities.
The third term replaces the cosmological constant of GR with another free function $P(X)$.
The fourth term represents all matter fields (collectively denoted by $\psi$), which couple minimally to the metric $g_{\mu \nu}$. 
$M \; (\ll M_P)$ is the EFT cut-off scale.

I will further focus on the subset of these theories in which $P(X)$ and $G_4 (X)$ are such that:
\begin{itemize} \vspace{-5pt}

\item[(i)] there is a stable background solution $\{ \bar{\phi} (t) , \bar{g}_{\mu\nu} (t) \}$ which coincides with that of $\Lambda$CDM on large scales (otherwise this theory would be immediately ruled out by observations),

\item[(ii)] the trivial configuration $\{ \phi , g_{\mu\nu} \}  = \{ 0 , \eta_{\mu\nu} \}$ is also a stable background solution (otherwise one cannot compute Lorentz-invariant amplitudes)$\,$\footnote{
This amounts to demanding that $P(X) = X + ...$ near $X=0$. 
},

\item[(iii)] the first two derivatives of $G_4^2 (X)$ may be comparable, but higher derivatives are suppressed on the cosmological background, i.e. $G_4^2 \sim  \bar{X} \partial_{X} G_4^2 \sim \bar{X}^2 \partial_X^2 G_4^2  \gg   \bar{X}^n \partial_X^n G_4^2$ for all $n \geq 3$.

\vspace{-5pt} \end{itemize}
\noindent While the background evolution can be made arbitrarily close to $\Lambda$CDM, this theory differs significantly in how linear perturbations evolve.
I will focus on two particular effects, parametrised by $c_T$ (the speed of gravitational waves relative to light$\,$\footnote{
Since~(\ref{eqn:Horndeski}) is an EFT for cosmological scales, it is not valid at $\geq 100$Hz. This $c_T$ is therefore the speed of very long-wavelength gravitational waves, and not the one probed by the LIGO/Virgo/KAGRA observatories.  
}) and $\alpha_B$ (the scalar/metric mixing, which affects large-scale clustering). 
These are determined by the first two derivatives of $G_4^2 (X)$, 
\bea
\left. \frac{ X \partial_X G_4^2 (X) }{G_4^2 (X) } \right|_{\bar{X} (t)}  = \frac{ c_T^2 - 1 }{c_T^2}  \, , \qquad
\left. \frac{ X^2 \partial_X^2 G_4^2 (X) }{G_4^2 (X) } \right|_{\bar{X} (t)}  =  \frac{ \alpha_B }{4 c_T^2}   - \frac{c_T^2 - 1 }{2 c_T^4}  \; . 
\label{eqn:cTaB_def}
\eea

Meanwhile, amplitudes for scattering $\phi$ fluctuations with both matter fields (e.g. the photon) and other $\phi$ fluctuations have recently been computed in this theory on the $\phi =0$ background,\cite{Melville:2019wyy,deRham:2021fpu}
\vspace{-16pt}
\bea
{A}_{\phi \gamma \to \phi \gamma} =&  \vcenter{\hbox{\includegraphics[scale=0.9]{{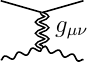}}}} \qquad \qquad \qquad \qquad\qquad\qquad\qquad   &\Rightarrow \quad c_{20} = \frac{2}{M_P M^3} \frac{ \partial_X G_{4}^2 (0) }{ G_4^2 (0) }  \, ,  \\
{A}_{\phi \phi \to \phi \phi} =&
\vcenter{\hbox{\includegraphics[scale=0.9]{{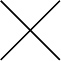}}}} \; + \; \vcenter{\hbox{\includegraphics[scale=0.9]{{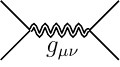}}}}  
\; + \; ( t, u \; \mathrm{channels} )
&\Rightarrow \quad c_{21} = - \frac{3}{M^6} \frac{ \partial_X^2 G_{4}^2 (0)}{G_4 (0) }  \; .  
\eea
Applying the UV/EFT correspondence (\ref{eqn:pos}), this EFT can only have a healthy UV completion if the first and second derivatives of $G_4^2$ have definite signs: $\partial_X G_4^2 (0) > 0$ and $\partial_X^2 G_4^2 (0) < 0$ (up to small corrections in $M/M_P$).
To implement these bounds in a cosmological setting, we invoke assumption (iii) and expand $G_4 (\bar{X}) = G_4 (0) + ...$ up to second order in $\bar{X}$ in~(\ref{eqn:cTaB_def}). This reveals,
\bea
\mbox{Causal/Unitary in UV} \;\; \Rightarrow \;\;  \alpha_B c_T^2 <  2 (c_T^2 - 1)  \;\; \mbox{and} \;\; \alpha_B c_T^2 <  2 (c_T^2 - 1) (2 c_T^2 + 1 ) \;\; \mbox{in IR.}
\label{eqn:pos_bounds}
\eea 

\begin{figure}
	\begin{minipage}{0.35\linewidth}
		\centerline{\includegraphics[width=0.9\linewidth]{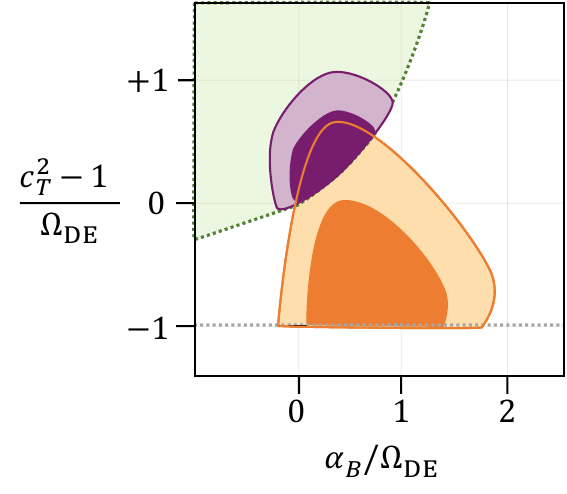}}
		\vspace{-0.2cm}
	\end{minipage}
	\hfill
	\begin{minipage}{0.60\linewidth}
	\vspace{-1cm}
	\begin{center}
		{\renewcommand{\arraystretch}{1.4}
			\setlength{\tabcolsep}{0.3cm}
			\begin{tabular}{| r | c c c|} \hline
				{} & $\alpha_B/\Omega_{\rm DE}$ & $\frac{ c_T^2 - 1}{\Omega_{\rm DE} }$ &  $\alpha_M/\Omega_{\rm DE}$ \\ \hline
				{\color[RGB]{220,135,42} no UV prior} & $0.71^{+0.90}_{-0.71}$ & $ -1^{+1.25}$ & $-0.02^{+1.32}_{-0.89}$     \\ \hline
				{\color[RGB]{120,28,109} with UV prior} & $0.26^{+0.46}_{-0.46}$ & $0.46^{+0.64}_{-0.41}$ &  $0.67^{+0.97}_{-0.58}$ \\ \hline
			\end{tabular}
		}
	\end{center}
	\end{minipage}
	\caption[]{Observational constraints on the $\{ \alpha_B, c_T \}$ parameter space.\cite{Melville:2019wyy} The positivity bounds (\ref{eqn:pos_bounds}) are satisfied in the green region. When implemented as a prior, they shift the posteriors from the orange to the purple intervals.}
	\label{fig:1}
\end{figure}

The simplest way to use these bounds is to first constrain the EFT parameter space using cosmological measurements,\footnote{In detail, we used a Markov chain Monte Carlo analysis with CMB temperature/lensing/polarisation (Planck 2015), baryon acoustic oscillations (SDSS/BOSS), matter power spectrum (SDSS DR4 LRG) and redshift space distortion (BOSS/6dF) to constrain the ratios $(c_T^2 - 1)/\Omega_{\rm DE}$, $\alpha_B / \Omega_{\rm DE}$ and $\alpha_M / \Omega_{\rm DE}$ (assumed constant in time).} and then compare the region favoured by data with the region~(\ref{eqn:pos_bounds})$\,$\footnote{
Since the dark energy density $\Omega_{\rm DE} \to 1$ at late times, this is when~(\ref{eqn:pos_bounds}) gives the constraint shown in Fig.~\ref{fig:1}.
}.
This is shown in Fig.~\ref{fig:1}. Surprisingly, the majority of the $68\%$ (and all of the $95\%$) confidence interval correspond to parameter values that are secretly unphysical: or at least, values that cannot be explained by this EFT unless the underlying UV theory violates causality or unitarity. 

Another approach is to limit the available parameter space using (\ref{eqn:pos_bounds}) as a theoretical prior, which after all are consequences of very basic assumptions (causality/unitarity).
In Fig.~\ref{fig:1} we compare the resulting posteriors with/without this UV prior$\,$\footnote{Note that in both cases we require that the EFT is stable and hence free of ghost and gradient instabilities for both the scalar and tensor: this imposes four restrictions on the parameters,\cite{deRham:2021fpu} including that $c_T^2 > 0$.}, and find that supplying this additional information about the UV improves the precision of our global fit by a factor of $\sim 3$. In particular, we also varied a third parameter, $\alpha_M$ (which quantifies $\partial_t \bar{X}$), and while this does not contribute to any amplitude constraint its marginalised distribution is affected by~(\ref{eqn:pos_bounds}), demonstrating that although UV priors from flat-space amplitudes are only sensitive to some properties of the EFT, they can have an important effect on all parameters in a global fit.

\paragraph{Conclusions.}
%
Using the fundamental axioms of causality and unitarity, I have shown that low-energy EFT coefficients can be expressed in terms of UV matrix elements that probe the underlying high-energy theory. This implies various ``positivity bounds'', which diagnose regions of EFT parameter space that can never be embedded into a UV-complete theory, and the phenomenological impact of these bounds is nicely illustrated by this simple EFT of dark energy.   

These bounds are necessary but not sufficient for a healthy UV completion. There are other constraints on this parameter space arising from the same assumptions of causality/unitarity, and so what has been presented here could certainly be made much stronger in the future. 

Also, the assumptions (ii) and (iii) are quite restrictive.\footnote{e.g. ghost-condensate models do not satisfy (ii), and the covariant Galileon cannot satisfy both (i) and (ii).}
One way to relax these assumptions is to consider scattering on a non-trivial background. Since some analogous amplitude constraints are known for backgrounds that break only boosts,\cite{Grall:2021xxm} positivity bounds have been derived for $\bar{\phi} (t) \propto t$ in shift-symmetric theories$\,$\cite{Davis:2021oce} and $\bar{\phi}(t) \propto t^2$ in Galileon-symmetric theories$\,$\cite{Melville:2022ykg}. In particular, if $G_4^2 (X)$ is dominated by a single $X^n$ term, then scattering on a $\bar{\phi} (t) \propto t$ background implies that $\partial_X^n G_4 (\bar{X}) < 0$ (except for $n=1$, for which $\partial_X G_4^2 > 0$).
This is evidence that the bound~(\ref{eqn:pos_bounds}) applies in a wider class of EFTs, but more work is certainly needed in this direction. 	 

Finally, the UV/EFT correspondence implies more than positivity conditions. High-energy information, such as which spins dominate these UV matrix elements, can now be reliably extracted from just one or two EFT coefficients. In this dark energy EFT, that amounts to a connection between the speed of low-energy gravitational waves and the spin of new small-scale physics. A refinement of Fig.~\ref{fig:1} would carve up the parameter space into regions with different high-energy properties, providing a useful map for future sky surveys. This is work in progress.

\paragraph{Acknowledgments.}
%
 S.M. is supported by a UKRI Stephen Hawking Fellowship (EP/T017481/1).

\paragraph{References.}


\vspace{-0.3cm}
\begin{thebibliography}{99}
\vspace{-0.2cm}
\bibitem{Adams:2006sv}A. Adams {\it et al}, \Journal{\it JHEP}{10}{014}{2006}.

\bibitem{deRham:2017avq}C. de Rham, S. Melville, A. J. Tolley and S.-Y. Zhou, \Journal{\PRD}{96}{081702(R)}{2017}.

\bibitem{Melville:2019wyy}S. Melville and J. Noller, \Journal{\PRD}{101}{021502}{2020}

\bibitem{deRham:2021fpu}C. de Rham, S. Melville and J. Noller, \Journal{\it JCAP}{08}{018}{2021}

\bibitem{Grall:2021xxm}T. Grall and S. Melville, \Journal{\PRD}{105}{L121301}{2022}

\bibitem{Davis:2021oce}A.-C. Davis and S. Melville, \Journal{\it JCAP}{11}{012}{2021}

\bibitem{Melville:2022ykg}S. Melville and J. Noller, \Journal{\it JCAP}{06}{031}{2022}

\end{thebibliography}
\end{document}